\documentclass[conference]{IEEEtran}

\usepackage{ifpdf}

\usepackage[dvips]{graphicx}
\graphicspath{{../eps/}}
\DeclareGraphicsExtensions{.eps}

\usepackage{mathbbol}
\usepackage{cite}
\usepackage[T1]{fontenc} 
\usepackage{algorithm}
\usepackage{algorithmic}
\usepackage[cmex10]{amsmath}
\usepackage{amsthm}
\usepackage{amssymb}
\usepackage{balance}
\usepackage{eqparbox}
\usepackage{textcomp}
\usepackage{ctable}
\usepackage{threeparttablex}
\usepackage[eulergreek]{sansmath}
\usepackage{color}
\usepackage[cmintegrals]{newtxmath}

\newtheorem{lemma}{Lemma}
\newtheorem{theorem}{Theorem}
\newtheorem{remark}{Remark}

\usepackage[open,openlevel=1]{bookmark}
\usepackage{bbold}
\usepackage{dsfont}
\usepackage{caption}

\allowdisplaybreaks

\makeatletter
\newcommand\fs@betterruled{%
  \def\@fs@cfont{\bfseries}\let\@fs@capt\floatc@ruled
  \def\@fs@pre{\vspace*{8pt}\hrule height.8pt depth0pt \kern2pt}%
  \def\@fs@post{\kern2pt\hrule\relax}%
  \def\@fs@mid{\kern2pt\hrule\kern2pt}%
  \let\@fs@iftopcapt\iftrue}
\floatstyle{betterruled}
\restylefloat{algorithm}
\makeatother

\newcommand{\bs}[1]{\boldsymbol{#1}}

\newcommand{\mb}[1]{\mathbf{#1}}
\newcommand{\mr}[1]{\mathrm{#1}}

\usepackage{scalerel}

\newcommand{\bseq}{\begin{subequations}}
	\newcommand{\eseq}{\end{subequations}}
\newcommand{\baln}{\begin{align}}
	\newcommand{\ealn}{\end{align}}
\newcommand{\balnd}{\begin{aligned}}
	\newcommand{\ealnd}{\end{aligned}}
\newcommand{\beq}{\begin{equation}}
	\newcommand{\eeq}{\end{equation}}
\newcommand{\beqn}{\begin{eqnarray}}
	\newcommand{\eeqn}{\end{eqnarray}}
\newcommand{\beqno}{\begin{eqnarray*}}
	\newcommand{\eeqno}{\end{eqnarray*}}
\newcommand{\bma}{\begin{displaymath}}
	\newcommand{\ema}{\end{displaymath}}
\newcommand{\bnu}{\begin{enumerate}}
	\newcommand{\enu}{\end{enumerate}}
\newcommand{\bce}{\begin{center}}
	\newcommand{\ece}{\end{center}}
\newcommand{\btb}{\begin{tabular}}
	\newcommand{\etb}{\end{tabular}}
\newcommand{\ba}{\begin{array}}
	\newcommand{\ea}{\end{array}}

\begin{document}
	
	\title{Energy-Efficient Precoding and Feeder-Link-Beam Matching Design for Bent-Pipe SATCOM Systems}
	\author{\IEEEauthorblockN{Vu Nguyen Ha, Juan Carlos Merlano Duncan, Eva Lagunas, Jorge Querol, and Symeon Chatzinotas} 
		
		\IEEEauthorblockA{\textit{Interdisciplinary Centre for Security, Reliability and Trust (SnT), University of Luxembourg, Luxembourg}}
 }
 \vspace{-5mm}




\IEEEcompsoctitleabstractindextext{
\begin{abstract}
	This paper proposes a joint optimization framework for energy-efficient linear precoding and feeder-link-beam matching design in a multi-gateway multi-beam bent-pipe satellite communication system. The proposed scheme jointly optimizes the precoding vectors at the gateway antennas and amplifying-and-matching mechanism at the satellite to maximize the system-weighted energy efficiency under the transmit power budget constraint. The technical designs are formulated into a non-convex sparsity problem consisting of a fractional-form objective function and sparsity-related constraints. To address these challenges, two iterative efficient designs are proposed by utilizing the concepts of Dinkelbach's method and the compressed-sensing approach. The simulation results demonstrate the effectiveness of the proposed scheme compared to another benchmark method.
\end{abstract}

}

\maketitle
\IEEEdisplaynotcompsoctitleabstractindextext
\IEEEpeerreviewmaketitle

\vspace{-1mm}

\section{Introduction}
\vspace{-2mm}
Recently, satellite communications (SATCOM) has been considered as an important component of the next generation of wireless communication which can enable seamless global connectivity. To meet the increasing high-data-rate demand, advanced satellite communication technologies have been developed for the traditional bent-pipe payload, including multi-gateway (GW) and multi-beam (MB) transmission \cite{fontanesi2023artificial,VuHa_ICCWks23}. 
Multiple GWs deployed in various areas can provide flexible and resilient connections between the ground segments and satellites \cite{VuHa_TWC23,VuHa_TVT23,fontanesi2023artificial,VuHa_ICCWks23,Eva_VTC22} while linear precoding (LP)-enabled transmission over MBs can mitigate the interference 
and improve the network performance significantly \cite{Tewel_TCom23,Liu_Access22, VuHaGC2022,VuHa_GCWks22, 9384476}. 
Regarding both user and feeder links (FLs), this advanced SATCOM system poses significant energy-efficient (EE) challenges, including the matching and amplifying issues at the payload. 

In recent years, several works have been proposed to optimize the EE of MB SATCOM systems. In \cite{Chatzinotas_Conf2011}, Chatzinotas et al. focused on investigating the EE of an MB downlink system using Minimum Mean Square Error (MMSE) LP and power optimization for the downlink channel. In \cite{Qi_WCL2020}, Qi et al. considered the design of EE multicast LP for multi-user MB SATCOMs under total power and Quality of Service (QoS) constraints. In \cite{Tedros_VTCFall2022}, Abdu et al. proposed an EE sparse LP design for SATCOM systems, where only a few LP coefficients are used with lower transmit power consumption depending on demand. Additionally, Joroughi et al. in \cite{Joroughi_TWC16} analyze the LP scheme in a multi-GW MB satellite system. The studied design is developed by utilizing a regularized singular value block decomposition of the channel matrix to minimize both inter-cluster and intra-cluster interference. 
These studies demonstrate the importance of EE and LP designs in SATCOMs; however, they have not considered the impact of the FLs in their optimization frameworks.

This paper considers an end-to-end forward link of a broadband multi-GW MB bent-pipe SATCOM system serving a number of ground users. In this scheme, the ground-based LP mechanism is assumed to precoded the user signals at the GWs. The precoded are then transmitted to the satellite through MIMO-enabled transmission from multiple GW antennas (GWAs) and sub-carriers (SCs). Then, the signals over different SCs, from different GWAs are matched to the transmitting antennas of different beams before being amplified and forwarded to the users. 
The system poses significant challenges for EE design due to the LP tasks, matching and amplifying mechanisms, and limited transmission-power budgets.
To address these challenges, we propose a joint optimization framework for EE-LP, FL-beam matching, and amplifying design. The proposed scheme optimizes the LP vectors at the GWs and sparsity variables regarding the forwarding process at the satellite to maximize the system EE (SEE) under the transmit power budget constraints. 
Dinkelbach's method and compressed-sensing approach are then employed to address the fractional-form objective function and sparsity-related critical issues.
The work provides two solutions balancing the overall power consumption and the Quality of Service (QoS) for all users. 
The simulation results are also presented to highlight the superior performance of the proposed approaches in comparison to another benchmark technique.

\section{System Model and Problem Formulation} \label{sec:SM}
\subsection{End-to-end Multi-GW Multi-beam SATCOM Systems} \label{ssec:E2ESys}
\vspace{-1mm}
\begin{figure}[!t]
\centering
\includegraphics[width=85mm,height=40mm]{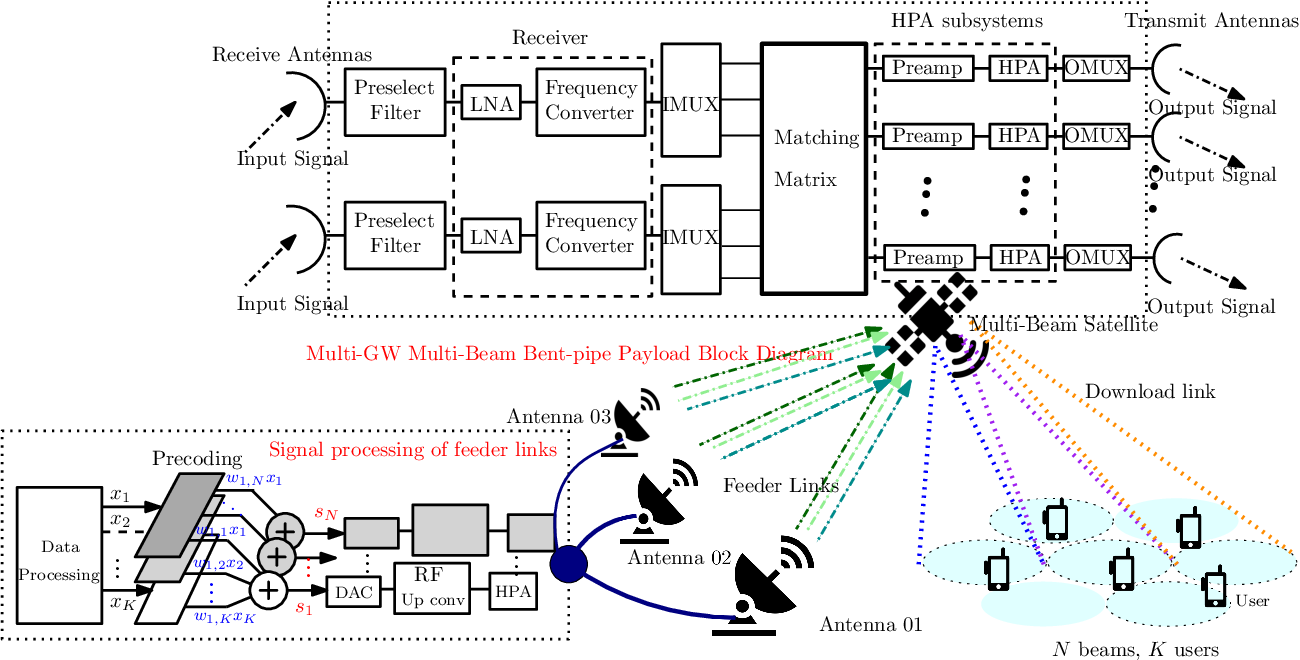}
\vspace{-2mm}
\captionsetup{font=footnotesize}
\caption{Transmission diagram of a multi-GW MB bent-pipe SATCOM system.}
\label{GEO_structure_fig}
\vspace{-4mm}
\end{figure}

Consider an end-to-end forward link of a broadband MB bent-pipe satellite system consisting of multiple GWs with $L$ distributed antennas on the ground, a bent-pipe transparent satellite (GEO - Geostationary Equatorial Orbit, MEO - Medium Earth Orbit, or LEO - Low Earth Orbit) equipped with $L$ receiving and $N$ transmission elements, and $K$ remote single-antenna users.
This scheme applies LP vectors to the corresponding symbol sequences for users at the GW antennas. 
Herein, the LP can be optimized centralizedly at the central controller. These precoded signals are sent to the satellite through the multiplexing transmission from multiple GWAs and SCs. The received signals are then amplified and forwarded to the users by the satellite payload.
\subsubsection{Multiplexing-enabled Feeder Links}
A multiplexing transmission is assumed for the communication between the GWAs and satellite with full re-use frequency of Q/V-band \cite{Delamotte_TAES20}.
Let $S$ be the number of SCs that are channelized at each GWA in its communication with the payload; hence, the FLs from $L$ GWAs can support at most $S L$ streams at a specific time\footnote{For instance, $4$GHz bandwidth over the Q/V band is channelized into $16$ $250$~MHz SCs. One SC carries one data stream without OFDM employed.}. 
Here, we assume\footnote{It required to note if $N > SL$, then the TDMA can be employed to transmit $N$ streams from GWAs to the satellite. And, if $N < SL$, one can select $N$ links to form a $N \times N$ FL channel matrix.} 
$N = S L$. 
At these higher frequencies, links to the satellite indeed use highly directional antennas such that strong LOS connections are established.
Following \cite{Delamotte_TAES20}, the FL channel matrix, i.e., $\mb{F}$, can be modelled as
$\mb{F} = \text{diag} \lbrace \mb{F}_{(1)}, \mb{F}_{(2)},...,\mb{F}_{(S)} \rbrace$.
Herein, $\mb{F}_{(s)} \in \mathbb{C}^{L \times L}$ represents the FL channel matrix of SC $s$ which can be expressed as
\vspace{-1mm}
\beq
\mb{F}_{(s)} = \scaleobj{.8}{\sqrt{G_{(s)}^{\sf{GWA}}G_{(s)}^{\sf{Sa-Rx}}}} \tilde{\mb{F}}_{(s)} \bs{\alpha}_{(s)},
\vspace{-1mm}
\eeq
where $G_{(s)}^{\sf{GWA}},G_{(s)}^{\sf{Sa-Rx}}$ are the antenna gains at GWAs and satellite, $\tilde{\mb{F}}_{(s)} \in \mathbb{C}^{L \times L}$ models the LOS free-space propagation.
Here, the $(m,n)$-entry of $\tilde{\mb{F}}_{(s)}$ is given by
$[\tilde{F}^{(s)}]_{(m,n)} = \sqrt{P^{\sf{Loss}}_{(m,n)}} \exp\lbrace -j (\psi_{m,n}+\phi_{m,n}) \rbrace$ where path-loss $P^{\sf{Loss}}_{(m,n)} = (c_0/4 \pi f_{(s)}^c r_{m,n})^2$, $\psi_{m,n} = {2 \pi f_{(s)}^c r_{m,n}}/{c_0}$ and $\phi_{m,n}$ represents the miss-synchronization phase noise. $\bs{\alpha}_{(s)} \in \mathbb{C}^{L \times L}$ is a diagonal matrix of the atmospheric impairments
experienced at the GWAs \cite{Delamotte_TAES20}.
The $l$-th diagonal element of $\bs{\alpha}_{(s)}$ can be given as
$\alpha^{(s)}_l = \vert \alpha_l \vert e^{-j \xi^{(s)}_l}$
where $\vert \alpha_l \vert \in (0,1]$ and $\xi^{(s)}_l \in [-\pi,\pi]$ are the amplitude fading and phase shift, respectively.

\subsubsection{User Links} Let $\mb{H} \in \mathbb{C}^{N \times K}$ be the channel matrix of the satellite-user links.
Herein, $[\mb{H}]_{(n,u)} = h_{n,u}$ stands for the channel coefficient from antenna $n$ to user $u$ which can be modeled using Rician channel model 
\cite{VuHaGC2022} as,
\vspace{-1mm}
\beq
h_{n,u} \!  = \! \sqrt{G^{\sf{gu}}_u P^{\sf{Loss}}_u} e^{-j \left(\psi_u+\phi_{n,u}\right)} \left( \scaleobj{.7}{\sqrt{\dfrac{\kappa}{\kappa+1}}} p^{\sf{pa}}_{n,u} +  \scaleobj{.7}{\sqrt{\dfrac{1}{\kappa+1}}}\alpha_{n,u} \right) ,
\vspace{-1mm}
\eeq
where $G^{\sf{gu}}_u$ is the user antenna receiving gain, path-loss $P^{\sf{Loss}}_u = (\lambda/4 \pi d_u)^2$; $\psi_u = \frac{2\pi d_u}{\lambda}$, $d_u$ is the distance between the satellite and user $u$, 
$p^{\sf{pa}}_{n,u}$ is the pattern coefficient of beam $n$ corresponding to the user's location; $\alpha_{n,u}$ is the small NLoS fading; $\kappa$ denotes Rician factor; $\lambda$ is the wavelength, and $\phi^{n,u}$ stands for the phase noise. 
\vspace{-1mm}
\begin{remark}
   Here, $\phi_{m,n}$ and $\phi_{n,u}$, are modeled as the summation of the phase noise caused by the imperfections from the hardware components, e.g., oscillators, of the GWAs, satellite payload and the users' receivers.
\end{remark}
\vspace{-1mm}
\subsubsection{Ground-based LP Design}
Due to $N$ transmission elements, one assumes the satellite can generate at most $N$ satellite beams for user-link transmission. 
Let $\mb{w}_u = [w_{1,u}, w_{2,u},...,w_{N,u}]^T \in \mathbb{C}^{N \times 1}$ be the LP vector designed for symbol sequence of user $u$, named $x_{u} \in \mathbb{C}$ and $\mathbb{E}_{x_{u}}\{|x_{u}|\} = 1$.
Considering the signal processing design, the GWs first apply all LP vector $\mb{w}_{u}$'s to the symbol sequences of all users. The precoded signals can be written as
$\mb{s} = \scaleobj{1}{\sum_{u \in \mathcal{U}}} \mb{w}_{u} x_u \in  \mathbb{C}^{N \times 1}$, where $\mathcal{U}$ stands for the set of users.
Then, $\mb{s}$ is sent to the satellite over $N$ FL SCs from $L$ GWAs.
The received signal at the satellite can be expressed as
\beq \label{eq:feed_received_signal}
\mb{r} = \mb{F}^{H} \mb{s} + \mb{n}^{\mr{fd}} = \mb{F}^{H} \scaleobj{.8}{\sum_{u \in \mathcal{U}}} \mb{w}_{u} x_u + \mb{n}^{\mr{fd}} ,
\eeq
where $\mb{n}^{\mr{fd}} \in \mathbb{C}^{N \times 1}$ is an AWGN vector at the satellite. 
\subsubsection{Payload Matching and Amplifying Process}
At the satellite, $N$ signal streams corresponding to $N$ FL SCs, i.e $\mb{r}$, are amplified and then matched to $N$ beams for propagation to users over the user links.
To ease the notation, we name FL~$t$ as the FL SC that carries the precoded symbol stream corresponding to $t$-th element of $\mb{s}$.
Let $\mb{B} \in \mathbb{R}_{+}^{N \times N}$ be the $N$-stream-to-$N$-beam amplifying and matching matrix. Denote $[\mb{B}]_{n,t} = b_{n,t}$ as an element locating on the $t$-th row and $n$-th column of $\mb{B}$, we have
$b_{n,t} > 0$ if $[\mb{r}]_t$ is transmitted over beam $n$, and $b_{n,t} = 0$ otherwise.
Due to the matching policy, $\mb{B}$ must be designed by regarding the following constraints,
\beq
(C1): \scaleobj{.8}{\sum_{\forall t}} \| b_{n,t} \|_0 \leq 1, \forall n, \text{ and } (C2): \scaleobj{.8}{\sum_{\forall n}} \| b_{n,t} \|_0 \leq 1, \forall t,
\eeq
where $\| x \|_0$ stands for the norm-$0$ of $x$. Multiplying $\mb{B}$ to $\mb{r}$ at the payload and then forwarding the amplified signal to users, one yields the received signal at all users as
\beq \label{eq:received_signal}
\mb{z} = \mb{H}^{H} \mb{B} \mb{r} + \mb{n}^{\mr{dl}} =  \mb{H}^{H} \mb{B} ( \mb{F}^{H} \mb{s} + \mb{n}^{\mr{fd}} )  + \mb{n}^{\mr{dl}},
\eeq
where $\mb{n}^{\mr{dl}} \in \mathbb{C}^{K \times 1}$ is an AWGN vector. 
Note that the $u$-th column of $\mb{H}$, i.e., $\mb{h}_u = [h_{1,u},h_{2,u},...,h_{N,u}]^T$, represents the channel vector from satellite to user $u$.  
Then, the received signal given in \eqref{eq:received_signal} yields the SINR at user $u$ as
\beq \label{SINR_user}
\scaleobj{.9}{\Gamma_u(\mb{W},\mb{B}) \! = \! \frac{ \left| \mb{h}^{H}_u \mb{B} \mb{F}^{H} \mb{w}_u \right|^2}{\sum_{i \neq u} \left| \mb{h}^{H}_{u} \mb{B} \mb{F}^{H} \mb{w}_i  \right|^2 \!\! + \!  \mb{h}^{H}_{u} \mb{B} \Sigma \mb{B}^T \mb{h}_{u} \! + \! {\sigma_u^{\mr{dl}}}^2}},
\eeq
where $\mb{W}=[\mb{w}_1, \mb{w}_2, ... \mb{w}_K] \in \mathbb{C}^{N \times K}$, ${\sigma_u^{\mr{dl}}}^2$  and $\Sigma = \mr{diag}[{\sigma_1^{\mr{fd}}}^2,{\sigma_2^{\mr{fd}}}^2,...,{\sigma_N^{\mr{fd}}}^2]$ represent the noise power at user $u$ and the noise covariance matrix at satellite, respectively. 
Learning from \eqref{SINR_user}, one can estimate the total achievable rate by the Shanon upper bound, as follows
\beq \label{eq:Shanon_rate}
R_{\sf{tot}}(\mb{W},\mb{B}) = R_s\scaleobj{.8}{\sum_{\forall u}} \log_2\left( 1 + \Gamma_u(\mb{W},\mb{B}) \right).
\eeq
where $R_s$ is the baud-rate of the user links. 
\begin{remark} \label{rmk1}
    Note that, 
    $\mb{B}=\mr{diag}(\bs{\xi})\mb{A}$ where $\mb{A}$ is a permutation matrix with $[\mb{A}]_{n,t} = a_{n,t} = \| b_{n,t} \|_0$ while $\xi_n$ is the amplify factor corresponding to beam $n$ and $\bs{\xi}=[\xi_1,...,\xi_N]^T$.
\end{remark}

\subsection{Power Consumption Model} \label{ssec:ECM}
\subsubsection{Gateway Power Consumption}
Besides the transmission power, the consumed component corresponding to the RF signal processing mainly depends on the number of FL subcarriers and activated beams.
Once, an FL SC is utilized, the corresponding precoded base-band signal goes through the DAC before being up-converted to the RF band, amplified by the HPA, and propagated to the satellite over that SC.
The transmission power relating to FL $t$ can be described as
$P^{\sf{GWA}}_t = \scaleobj{1}{\sum_{\forall u}} \mb{w}^H_{u}\mb{E}_t\mb{w}_{u}$,
where $\mb{E}_t$ is a diagonal matrix in $\mathbb{R}^{N \times N}$ with zero elements and one at the $t$-th position.
Note that FL $t$ is utilized if and only if $P^{\sf{GWA}}_t >0$. Then, the number of utilized FLs can be described as
$T_{\mr{fd}} = \sum_{\forall t}  \| P^{\sf{GWA}}_t \|_0$.
Moreover, the power consumption of HPA for FL transmission can be modeled as $P^{\mr{PA}}_{{\sf{GWA}},t} = (1/\rho_{\sf{GWA}}) (P^{\sf{GWA}}_t - P_{\mr{bb}})$, \cite{VuHa_TGCN20}
in which $\rho_{\sf{GWA}}$ stands for the HPA efficiency and $P_{\mr{bb}}$ is the base-band signal power.
Then, the RF signal processing and propagation power consumption of all GWAs is
\beqn
P_{\sf{GWA}}(\mb{W}) & = &  P^{\mr{hw}}_{\sf{GWA}} T_{\mr{fd}}  + \scaleobj{.8}{\sum \limits_{\forall t}} \left( P^{\mr{PA}}_{{\sf{GWA}},t} + P^{\sf{GWA}}_t \right)   \\
& = & P^{\mr{hw}}_{\sf{GWA}}  \scaleobj{.8}{\sum_{\forall t}} \| P^{\sf{GWA}}_t \|_0  + \scaleobj{.7}{\dfrac{\rho_{\sf{GWA}}+1}{\rho_{\sf{GWA}}} \sum \limits_{u \in \mathcal{U}}} \mb{w}_u^H \mb{w}_u, \nonumber
\eeqn
where $P^{\mr{hw}}_{\sf{GWA}}$ is the total power of DAC, RF up-converter components, and $-P_{\mr{bb}}/\rho_{\sf{GWA}}$. Here, we assume that $P^{\mr{hw}}_{\sf{GWA}} > 0$.

\subsubsection{Satellite Power Consumption} 
According to the bent-pipe transponder illustrated in Fig.~\ref{GEO_structure_fig}, the satellite power consumption can be estimated as
\beqn
&& \hspace{-1cm} P_{\mr{Sa}}(\mb{W},\mb{B})  =  P^{\mr{hw}}_{\mr{Sat}} T_{\mr{fd}}  +  P^{\mr{PA}}_{\mr{Sa}} + P^{\mr{dl}}_{\mr{Tx}}  \nonumber \\
&& \hspace{-1cm} = P^{\mr{hw}}_{\mr{Sat}} \! \scaleobj{.8}{\sum_{\forall (t,n)}} \! \| P^{\sf{GWA}}_t \|_0  \! + \! \scaleobj{.8}{\dfrac{\rho_{\mr{Sa}}+1}{\rho_{\mr{Sa}}}}  (  \scaleobj{.8}{\sum \limits_{u \in \mathcal{U}}} \mb{w}_u^H \mb{F} \mb{B}^T \mb{B} \mb{F}^H \mb{w}_u \! + \!\mr{Tr}( \mb{B}\Sigma\mb{B}^T) ) \nonumber \\
&& \hspace{2cm} -(1/\rho_{\mr{Sa}})(\scaleobj{.8}{\sum \limits_{u \in \mathcal{U}}} \mb{w}_u^H \mb{F} \mb{F}^H \mb{w}_u \! + \!\mr{Tr}( \Sigma) ),
\eeqn
where $P^{\mr{hw}}_{\mr{Sat}}$ is for the power of satellite hardware components, $P^{\mr{dl}}_{\mr{Tx}}$ is the transmission power and $P^{\mr{PA}}_{\mr{Sa}} = (1/\rho_{\mr{Sa}})(|\mb{B} \mb{r}|^2 - |\mb{r}|^2)$ implies the HPA power in which $\rho_{\mr{Sa}}$ is the power amplifier efficiency at the satellite \cite{VuHa_TGCN20}.
\subsubsection{Total Weighted Power Consumption}
From the engineering point of view, we aim to utilize various weights for power consumption from GWAs and satellite due to the different energy budgets of these system components. In particular, a higher weight should be 
marked for satellite due to its limited power-supply sources.
Let $\delta^{\sf{GWA}}$ and $\delta^{\sf{Sa}}$ be the impacting weights corresponding to the power consumption of GWAs and satellite, respectively.
Then, the total weighted power consumption can be expressed as
\beqn
P_{\mr{tot}}(\mb{W},\mb{B})  = \delta^{\sf{GWA}} P_{\sf{GWA}}(\mb{W}) + \delta^{\sf{Sa}} P_{\mr{Sa}}(\mb{W},\mb{B}).  \label{eq:P_tot}
\eeqn

\subsection{Problem Formulation}
We are now ready to define the ratio of the sum rate to the total weighted power consumption, so-called system weighted energy efficiency (SWEE) in bits/W, as
\beq
\eta(\mb{W},\mb{B})={R_{\sf{tot}}(\mb{W},\mb{B})}/{P_{\mr{tot}}(\mb{W},\mb{B})}.
\eeq
In this paper, we are interested in jointly optimizing the LP vectors at the GWs, and the matching and amplifying gains at the satellite to maximize the SWEE under the constraint on the transmit power budget at each antenna. This SWEE maximization (SWEEM) problem can be stated as
\begin{subequations} \label{SEE-max-prb}
\begin{eqnarray} 
	\hspace{-1cm}&\underset{\mb{W},\mb{B}}{\max}& \hspace{-0.2cm} \eta(\mb{W},\mb{B})={R_{\sf{tot}}(\mb{W},\mb{B})}/{P_{\mr{tot}}(\mb{W},\mb{B})} \label{obj_func_SEE}\\
	\hspace{-1cm}&\text{s. t. }&  \hspace{-0.2cm}\text{constraints }(C1), (C2), \nonumber \\
	\hspace{-1cm}&& \hspace{-0.2cm} (C3): \scaleobj{.8}{\sum_{\forall u}} \mb{w}^H_{u}\mb{E}_t\mb{w}_{u} \leq \bar{P}^{\sf{GWA}}_t, \forall t, \label{cnt2} \\
	\hspace{-1cm}&& \hspace{-0.5cm} (C4): \scaleobj{.8}{\sum_{\forall t}} b^2_{n,t} ( \scaleobj{.8}{\sum \limits_{u \in \mathcal{U}}} \mb{w}_u^H \mb{F} \mb{E}_t \mb{F}^{H} \mb{w}_u +{\sigma^{\mr{fd}}_t}^2 )  \leq \bar{P}^{\mr{Sa}}_n, \forall n, \label{cnt3}
\end{eqnarray}
\end{subequations}
where $(C3)$ and $(C4)$ are considered based on the transmission power budget of each FL at the GWs and every antenna of the satellite, respectively.
As can be seen, problem (\ref{SEE-max-prb}) is an NP-hard mixed integer programming.
To deal with this complicated problem, we aim to employ Dinkelbach's method \cite{Dinkelbach67} 
and compressed-sensing approach to cope with 
the fractional-form critical issue and mixed-integer challenge.


\section{Proposed Solution Approaches}
\label{sec2}
\subsection{The Foundation of Dinkelbach Method}
This method is summarized in the following theorem \cite{Dinkelbach67}. 

\begin{theorem} \label{thr_eta_star}
Let $\eta^{*}$ be the optimal objective value of problem $(\mathcal{P}_I): \underset{\mb{x}}{\max} \; R(\mb{x})/P(\mb{x}) \; \text{s. t. } \mb{x} \in \mathcal{S}$, where $P(\mb{x}) > 0$ $\forall \mb{x} \in \mathcal{S}$. Consider the subtracting-form problem $
\left( \mathcal{P}_{II}^{(\eta)} \right):  \underset{\mb{x}}{\max} \; R(\mb{x}) - \eta P(\mb{x}) \; \text{s. t. } \mb{x} \in \mathcal{S}$.
Denote $\chi(\eta)$ as the optimal objective value of $\left( \mathcal{P}_{II}^{(\eta)} \right)$ for given $\eta$. Then, $\chi(\eta)$ is a function of $\eta$ which has the following characteristics:
\begin{itemize}
	\item[i)] $\chi(\eta)$ is a strictly monotonic decreasing function.
	\item[ii)] $\chi(\eta) > 0$ if and only if $\eta < \eta^{*}$, vice versa.
	\item[iii)] $(\mathcal{P}_I)$ and $\left( \mathcal{P}_{II}^{(\eta^{*})} \right)$ have the same set of optimal solutions. 
\end{itemize}
\end{theorem}
\begin{IEEEproof}
The proof can be found in \cite{Dinkelbach67,VuHa_TGCN20}.
\end{IEEEproof} 
Theorem~\ref{thr_eta_star} prompts us to develop an iterative approach to obtain the optimal solution of problem \eqref{SEE-max-prb} which is summarized in Algorithm~\ref{P2_alg:1}.
In particular, we first state the parameterized problem
for a given value of $\eta$ as follows.
\beq
\underset{\mb{W},\mb{B}}{\max} \; R_{\sf{tot}}(\mb{W},\mb{B})- \eta P_{\mr{tot}}(\mb{W},\mb{B}) \text{ s.t. } (C1)-(C4).
\label{sub-tract-prob}
\eeq
Then, the algorithm tends to iteratively solve problem \eqref{sub-tract-prob} for a certain value of $\eta$, and adjust $\eta$ until an optimal $\eta^{\star} \geq 0$ satisfying $ R_{\sf{tot}}(\mb{W},\mb{B})- \eta^{\star} P_{\mr{tot}}(\mb{W},\mb{B})=0$ is found. In what follows, we will propose two novel approaches dealing with problem \eqref{sub-tract-prob} in \textbf{Step 3} of Algorithm~\ref{P2_alg:1} efficiently.


\begin{algorithm}[!t]
\caption{\footnotesize \textsc{Overview of the Proposed Algorithm}}
\label{P2_alg:1}
\footnotesize
\begin{algorithmic}[1]
	\STATE Initialize $\eta^{(0)}=0$, set $\ell=0$, and choose a tolerate $\tau^{\mr{out}}$.
	\REPEAT 
	\STATE Solve \eqref{sub-tract-prob} with $\eta^{(\ell)}$ to achieve $(\mb{W}^{(\ell)},\mb{B}^{(\ell)})$.
	\STATE Update $\eta^{(\ell+1)}=\frac{R_{\sf{tot}}(\mb{W}^{(\ell)},\mb{B}^{(\ell)})}{P_{\mr{tot}}(\mb{W}^{(\ell)},\mb{B}^{(\ell)})}$.
	\STATE Set $\ell:=\ell+1$.
	\UNTIL $\vert \eta^{(\ell)} - \eta^{(\ell-1)} \vert \leq \tau^{\mr{out}}$.
	\STATE Return $(\mb{W}^{(\ell-1)},\mb{B}^{(\ell-1)})$.
\end{algorithmic}
\normalsize
\end{algorithm}


\subsection{Joint Linear Precoding and FL-Beam Matching Design}
The challenges of solving problem \eqref{sub-tract-prob} come from the non-convex objective function and the sparsity terms. To address these, the MMSE-based transformation \cite{VuHaGC2022} and the CS method can be exploited as follows.
\subsubsection{MMSE-based Transformation}
We first relate the logarithm-formed rate to a weighted sum-mean square error (MSE) minimization problem in the following theorem.
\begin{theorem}
\label{P2_thr2}
Problem \eqref{sub-tract-prob} is equivalent to the following,
\vspace{-1mm}
\beq
\underset{ \mb{W},\mb{B}}{\min}\;
\eta P_{\mr{tot}}(\mb{W},\mb{B})  + R_s \!\scaleobj{.7}{\sum \limits_{\forall u}} \!\!\left(\omega_{u} e_u \! - \!  \log \omega_{u}\right)  \text{ s.t. } (C1)-(C4), \label{prob:neurody03}
\vspace{-2mm}
\eeq
where $e_u = \mathbb{E} [ |x_{u} - \delta_{u} z_{u}|^2]$, $\omega_{u}$ and $\delta_{u}$ represent the MSE weight and the receive coefficient for user $u$, respectively.
\end{theorem}
\begin{IEEEproof} 
The proof is similar to that given in \cite{VuHaGC2022}. 
\end{IEEEproof}

It is noted that problem \eqref{prob:neurody03} is not \emph{jointly} convex, it is convex over each set of variables $\mb{W}$, $b_{n,t}$'s, $\delta_{u}$'s, and $\omega_{u}$'s.
Thus, one can solve problem \eqref{prob:neurody03} by alternately optimizing over one set of variables
while keeping the others fixed.

\subsubsection{Update MSE Weights and Receive Coefficients}
Handling some minor manipulation on $e_u = \mathbb{E} [ |x_{u} - \delta_{u} z_{u}|^2]$ and taking the corresponding derivative, $\delta_{u}$'s can be optimized in order to minimize $e_u$ for given $(\mb{W},\mb{B})$ as
\vspace{-1mm}
\begin{eqnarray} \label{receive2}
\delta_{u}^{\star} =  \bs{\Theta}_u^{-1}\mb{w}_u^H \mb{F}\mb{B}^T \mb{h}_u, 
\end{eqnarray}
where $\bs{\Theta}_u = \sum \limits_{\forall i} \left| \mb{h}^{H}_{u} \mb{B} \mb{F}^{H} \mb{w}_i  \right|^2 \!\! + \!  \mb{h}^{H}_{u} \mb{B} \Sigma \mb{B}^T \mb{h}_{u} \! + \! {\sigma_u^{\mr{dl}}}^2$.
Again, by taking the derivative of the objective function in \eqref{prob:neurody03} with respective to $\omega_u$, the optimum value $\omega_u^{\star}$ can be expressed as
\vspace{-1mm}
\beqn \label{omega2}
\omega_u^{\star} \! = \! e_u^{-1} \! = \! \left(1 \! - \bs{\Theta}_u^{-1}\left| \mb{h}^{H}_{u} \mb{B} \mb{F}^{H} \mb{w}_u  \right|^2\right)^{-1}.
\eeqn

\subsubsection{Linear Precoding and Amplifying Matrix Design}
We are now ready to develop an efficient mechanism to due with \eqref{prob:neurody03} for given $\delta_{u}$'s and $\omega_u$'s.
As can be observed, the challenges of solving $\mb{W}$ and $\mb{B}$ come from the norm-$\ell_0$ forms of both of these variables in the power consumption formulas and constraints $(C1)-(C2)$.
To simply such difficulty corresponding $\mb{W}$, we transform the term $\| P^{\sf{GWA}}_t \|_0$ into the sparsity form of $\mb{B}$ by regarding the following lemma.
\begin{lemma} \label{lm01}
Regarding the optimal solutions of problems \eqref{SEE-max-prb} (also \eqref{sub-tract-prob} and \eqref{prob:neurody03}), the following equality can be hold,
\beq
\| P^{\sf{GWA}}_t \|_0 = \Vert \scaleobj{0.7}{\sum_{\forall n}} b_{n,t} \Vert_0 = \scaleobj{0.7}{\sum_{\forall n}} \Vert b_{n,t} \Vert_0, \quad \forall t.
\eeq
\end{lemma}
\begin{IEEEproof}
As can be seen, if $P^{\sf{GWA}}_t = 0$ which implies that FL $t$ is not activated; then, $b_{n,t} = 0$ for all $n$ can be an efficient solution. 
Inversely, $\sum_{\forall n} b_{n,t} = 0$ shows that no beam will forward the signal from FL $t$ to users. In such scenarios, to achieve better solutions, $P^{\sf{GWA}}_t$ must be zeros. 
\end{IEEEproof}
Thanks to Lemma~\ref{lm01} and regrading that $e_u  = [ 1 + \left| \delta_u \right|^2 \bs{\Theta}_u - 2 \Re \left(  \delta_u^{\prime}\mb{w}_u^{H} \mb{F}\mb{B}^T \mb{h}_{u} \right)]$, one can rewrite problem \eqref{sub-tract-prob} for given $\delta_{u}$'s and $\omega_u$'s as
\vspace{-1mm}
\begin{eqnarray} 
\hspace{-0.4cm}&\underset{\mb{W},\mb{B}}{\min}& \hspace{-0.2cm}
\! \mr{Tr}((\nu_3\mb{I} +\Lambda)\mb{B}\Sigma\mb{B}^T ) \! - \!2\!  \scaleobj{.7}{\sum \limits_{u \in \mathcal{U}}} \! \Re (  \omega_u \delta_u^{\prime}\mb{w}_u^{H} \mb{F}\mb{B}^T \mb{h}_{u} ) \nonumber \\
\hspace{-0.4cm} & &  \hspace{-0.7cm}  + \!
\scaleobj{.7}{\sum \limits_{\forall u}} \mb{w}_u^H \! [  \nu_1 \mb{I} \! - \! \nu_2\mb{F}\mb{F}^H \!\! + \! \mb{F} \mb{B}^T \! ( \nu_3 \mb{I}+ \Lambda) \mb{B} \mb{F}^H ]  \mb{w}_u \! + \! \nu^{\sf{hw}} \!\! \scaleobj{.7}{\sum_{\forall(t,n)}} \!\! \| b_{n,t} \|_0  ,
\label{obj_func_SEE} \nonumber \\
\hspace{-0.4cm}&\text{s. t. }&  \hspace{-0.2cm}\text{constraints }(C1)-(C4), \label{min-WB} 
\vspace{-1mm}
\end{eqnarray}
where $\Re(.)$ stands for the real part, $\nu^{\sf{hw}} = \scaleobj{.7}{\dfrac{\eta}{R_s}}( \delta^{\sf{GWA}}P^{\mr{hw}}_{\sf{GWA}} + \delta^{\sf{Sa}}P^{\mr{hw}}_{\mr{Sa}}) $, $\nu_1 = \scaleobj{.7}{\dfrac{\eta \delta^{\sf{GWA}}}{R_s}} \scaleobj{.7}{\dfrac{\rho_{\sf{GWA}}+1}{\rho_{\sf{GWA}}}}$, 
$\nu_2 = \scaleobj{.7}{\dfrac{\eta\delta^{\mr{Sa}}}{\rho_{\mr{Sa}}R_s}}$, $\nu_3 = \scaleobj{.7}{\dfrac{\eta \delta^{\sf{Sa}}}{R_s}} \scaleobj{.7}{\dfrac{\rho_{\mr{Sa}}+1}{\rho_{\mr{Sa}}}}$, and
$\Lambda = \sum_{\forall i} \omega_i \left| \delta_i \right|^2 \mb{h}_i \mb{h}_i^H$.
\paragraph{Linear Precoding Design}
For given $\mb{B}$, the corresponding LP vectors can be determined by solving the following Quadratically Constrained Quadratic
Program (QCQP),
\beq
\underset{\mb{W}}{\min} 	\scaleobj{0.7}{\sum \limits_{\forall u}}  \mb{w}_{u}^H \bs{\Pi}  \mb{w}_u  - 2 \Re (  \mb{w}_u^{H} \mb{k}_{u} )  \text{ s.t. } (C3) \text{ and } (C4), \label{QCQP-probW}
\vspace{-1mm}
\eeq
where $\bs{\Pi} \!=\! \nu_1 \mb{I}\! -\! \nu_2\mb{F}\mb{F}^H \!+\!  \mb{F} \mb{B}^T \!( \nu_3 \mb{I}\!+\! \Lambda) \mb{B} \mb{F}^H$ and $\mb{k}_{u} = \omega_u\delta_u^{\prime} \mb{F}\mb{B}^T \mb{h}_{u} $. This QCQP problem can be solved effectively by employing some standard convex optimization solvers.
\paragraph{Sparsity Amplifying Matrix Design}
To deal with the norm-$\ell_0$ challenge of solving $\mb{B}$, one can employ the re-weighted norm-$\ell_1$ approximation methods which have been proposed
to enhance the data acquisition in compressed sensing. In
particular, the sparsity term $\| b_{n,t} \|_0$ can be approximated to $\beta_{n,t} b_{n,t}$ where $\beta_{n,t}$ is a re-weighted factor. In the CS approach, a such factor can be chosen as \cite{VuHa_TWC18,VuHa_ICC18,VuHa_TWC21}
\beq \label{beta_update}
\beta_{n,t} = \scaleobj{0.8}{\sqrt{1/(b_{n,t}^2 + \epsilon)}},
\vspace{-1mm}
\eeq
where $\epsilon \ll 1$. Note that $\beta_{n,t}$ can be updated so that the closed-to-zero elements in the previous iteration will suffer a huge penalty.
Denote $\mb{b}_t \in \mathbb{C}^{N \times 1}$ as the vector generated from the $t$-th column of $\mb{B}$. 
Regarding that $\| b_{n,t} \|_0 = \| b_{n,t}^2 \|_0$, we can rewrite the sparsity terms in \eqref{min-WB} as
\beq
\scaleobj{0.7}{\sum_{\forall n}} \| b_{n,t} \|_0 = \mb{b}_t^T \mb{D}_t  \mb{b}_t \text{ and } \scaleobj{0.7}{\sum_{\forall t}} \| b_{n,t} \|_0 = \scaleobj{0.7}{\sum_{\forall t}} \mb{b}_t^T \mb{E}_{n,t} \mb{b}_t,
\eeq
where $\mb{D}_t = {\sf{Diag}}(\beta^2_{1,t}, ..., \beta^2_{N,t})$ and $\mb{E}_{n,t}$ is a zero matrix except that its $n$-th diagonal element is $\beta^2_{n,t}$.
Then, we introduce vector $\mb{b} \in \mathbb{C}^{N^2 \times 1}$ which is $\mb{b}=[\mb{b}_1;...;\mb{b}_N]$.
By properly choosing and updating $\beta_{n,t}$’s,
problem \eqref{min-WB} for given $\mb{W}$ can be relaxed to the following QCQP problem,
\begin{subequations} \label{B_prob}
\begin{eqnarray} 
	\hspace{-1cm} \underset{\mb{B}}{\min} \,\mb{b}^T \!\!(\bs{\Psi} \!+ \!\nu^{\sf{hw}}\mb{D}\!+\!\nu_3\tilde{\bs{\Sigma}}) \mb{b} \! - \! \tilde{\mb{f}}^T\mb{b} 
	& \hspace{-0.4cm}  \text{ s.t. }& \hspace{-0.4cm}  (\tilde{C}1)\!: \mb{b}^T \! \tilde{\mb{D}}_t  \mb{b} \! \leq \! 1, \forall t, \\
	\hspace{-1cm}&& \hspace{-5cm} (\tilde{C}2):  \mb{b}^T \mb{E}_n \mb{b} \leq 1, \forall n, \text{ and } (\tilde{C}4):  \mb{b}^T \bs{\Gamma}_n \mb{b} \leq \bar{P}^{\mr{Sa}}_n, \forall n, 
\end{eqnarray}
\end{subequations}
where $\mb{D}={\sf{BlkDiag}}(\mb{D}_1;...,\mb{D}_N)$; $\tilde{\bs{\Sigma}} ={\sf{BlkDiag}}({\sigma_1^{\mr{fd}}}^2\mb{I},...,$ ${\sigma_N^{\mr{fd}}}^2\mb{I})$;  $\bs{\Psi} \in \mathbb{C}^{N^2 \times N^2}$ and its $(n,t)$-th $N \times N$ block matrix is defined as
$[\bs{\Psi}]_{(n,t)} = (\nu_3 \mb{I} + \scaleobj{0.9}{\sum_{\forall u}}\omega_u |\delta_u|^2 \mb{h}_u \mb{h}_u^H)(\scaleobj{0.9}{\sum_{\forall i}}\mb{w}_i^H \mb{f}_n\mb{f}_t^H \mb{w}_i)$; $\tilde{\mb{f}} \! \in \! \mathbb{R}^{N^2 \times 1}$ and its $t$-th $N \!\times \! 1$ block vector is defined as $[\tilde{\mb{f}}]_t$ $=2\Re( \scaleobj{0.9}{\sum_{\forall u}}\omega_u \delta^{\prime}_u\mb{h}_u\mb{w}_u^H\mb{f}_t)$; $\tilde{\mb{D}}_t \in \mathbb{C}^{N^2 \times N^2}$ contains all zeros except that its $(t,t)$-th $N \times N$ block matrix is $\mb{D}_t$; $\mb{E}_n =  {\sf{BlkDiag}}(\mb{E}_{n,1};...;\mb{E}_{n,N})$; and $\bs{\Gamma}_n = {\sf{BlkDiag}}(\gamma_1 \mb{E}_{n,1};...; $ $\gamma_N \mb{E}_{n,N})$ in which $\gamma_t = \scaleobj{1}{\sum_{\forall u}} \mb{w}_u^H \mb{f}_t \mb{f}_t^H \mb{w}_u +{\sigma^{\mr{fd}}_t}^2$. Herein, $\mb{f}_t$ represents the vector generated from $t$-th column of $\mb{F}$.
This QCQP problem can also be solved by employing some \textit{off-the-shelf} convex
optimization solvers.
\subsubsection{Joint LP and FL-Beam Matching Algorithm}
By iteratively updating $\beta_{n,t}$'s and alternatively determining $\omega_u$'s, $\delta_u$'s, $\mb{W}$, $\mb{B}$ as described above, the solution of \eqref{sub-tract-prob} can be obtained. The solution approach is summarized in Algorithm~\ref{P2_alg:3}. Then, the system energy efficiency (SEE) results can be obtained by integrating Algorithms ~\ref{P2_alg:1} and \ref{P2_alg:3}, which is also named as \textit{joint LP and FL-beam matching} (JPFBM) mechanism.
\begin{remark}
   Giving the convergence proof of this algorithm is challenging
due to the lack of space. However, convergence can be guarantee by Dinkelbach Method, the well-known compressed-sensing approach, as well as the goal of updating $\{\delta_{u},\omega_u\}$'s,  $\mb{W}$, and $\mb{B}$ which aims to keep the objective of problem \eqref{prob:neurody03} monotonically decreasing.
\end{remark}

\vspace{-3mm}

\begin{algorithm}[!t]
\caption{\footnotesize \textsc{Iterative Joint LP and FL-Beam Matching Design}}
\label{P2_alg:3}
\footnotesize
\begin{algorithmic}[1]
	\STATE Initialize: Select suitable $\mb{W}^{[0]}$, and small $\epsilon$ and set $\mb{B}^{[0]} = \vmathbb{1}_{N \times N}$. Set $k=0$. 
	\REPEAT 
	\STATE Update $k:=k+1$.
	\STATE Calculate $\{\delta_{u}^{[k]},\omega_u^{[k]}\}$'s as in \eqref{receive2}, \eqref{omega2} based on  $\mb{W}^{[k-1]}$, $\mb{B}^{[k-1]}$.
	\STATE Optimize $\mb{W}^{[k]}$ by solving problem \eqref{QCQP-probW}.
	\STATE Update $\beta_{n,t}$'s as in \eqref{beta_update}.
	\STATE Optimize $\mb{B}$ by solving problem \eqref{B_prob}.
	\UNTIL Convergence of the objective function in \eqref{prob:neurody03}.
\end{algorithmic}
\normalsize
\end{algorithm}

\subsection{Low-Complex Solution Approach for given Matching}
Thanks to Remark~\ref{rmk1}, we aim to optimize the LP and amplifying designs for a given FL-beam matching solution.
Note that the corresponding amplifier gain should be set to zero when FL $t$ is inactivated, which yields $\| P^{\sf{GWA}}_t \|_0 = \| \xi_t \|_0$. 
Re-employing the compressed-sensing approach for treating variables $\bs{\xi}$, we introduce the re-weight factor $\alpha_t$ as 
$\alpha_t = \scaleobj{.8}{\sqrt{1/(\xi_t^2 + \epsilon)}}$.
Then, for given $\mb{A}$, problem \eqref{B_prob} can be stated as
\vspace{-1mm}
\beq 
	\underset{ \bs{\xi}}{\min} \;
	\bs{\xi}^T (\bs{\Phi}+\nu^{\sf{hw}} \mb{L}) \bs{\xi}  - \mb{c}^T \bs{\xi} \text{ s.t. } \xi^2_n \scaleobj{.7}{\sum_{\forall t}} a_{n,t} \gamma_t  \leq \bar{P}^{\mr{Sa}}_n, \forall n, \label{prob:XiY2}
 \vspace{-2mm}
\eeq
where $\bs{\Phi} \in \mathbb{C}^{N \times N}$ and its $(n,t)$-th elements is defined as $\bs{\Phi}_{(n,t)}= \mb{a}_n^T [\bs{\Psi}]_{(n,t)} \mb{a}_t$; $\mb{L} = {\sf{Diag}}(\alpha^2_1, ..., \alpha^2_N)$; $\mb{c} \in \mathbb{R}^{N \times 1}$ and its $t$-th element is defined as $c_t = \mb{a}_t^T [\tilde{\mb{f}}]_t$. 
This problem is also a QCQP where $\bs{\xi} \in \mathbb{R}^{N \times 1}$; hence, $\bs{\xi}$ can be defined optimally by employing some \textit{off-the-shelf} optimization tools. Then, the proposed continuous-rate AF precoding design framework is summarized in Algorithm~\ref{P2_alg:4}. Then, the SEE results are obtained by integrating Algorithms ~\ref{P2_alg:1} and \ref{P2_alg:4}, which is also called \textit{joint LP with amplify-and-forward} (JPAF) approach.

\begin{algorithm}[!t]
\caption{\footnotesize \textsc{Joint LP with Amplify-and-Forward Design}}
\label{P2_alg:4}
\footnotesize
\begin{algorithmic}[1]
	\STATE Initialize: 
 \begin{itemize}
 \item Define a matching matrix $\mb{A}$ satisfying $(C1)$ and $(C2)$.
 \item Select suitable $\mb{W}^{[0]}$, and small $\epsilon$, and set $\bs{\xi}^{[0]} = \vmathbb{1}_{N \times 1}$. Set $k=0$.
 \end{itemize}
    \REPEAT 
	\STATE Update $k:=k+1$.
	\STATE Define $\{\delta_{u}^{[k]},\omega_u^{[k]}\}$'s as in \eqref{receive2}, \eqref{omega2} based on  $\mb{W}^{[k-1]}$, $\mb{A}$, $\bs{\xi}^{[k-1]}$.
	\STATE Optimize $\mb{W}^{[k]}$ by solving problem \eqref{QCQP-probW}.
	\STATE Update $\alpha_t$'s as $\alpha_t = \scaleobj{.8}{\sqrt{1/(\xi_t^2 + \epsilon)}}$, Optimize $\bs{\xi}$ by solving problem \eqref{prob:XiY2}.
	\UNTIL Convergence of the objective function in \eqref{prob:neurody03}.
\end{algorithmic}
\normalsize
\end{algorithm}

\vspace{-1mm}

\section{Simulation Results}

\vspace{-1mm}
\begin{table}[!t]
			\centering
   \captionsetup{font=footnotesize}
		\caption{Simulation Parameters}
		\label{tab:simpara}
		\vspace{-0.2cm}
		\begin{tabular}{l | r }
			\toprule
			\midrule
            GW Hardware-Power & $10$ W \\
            $[49.075, 49.325, 49.575, 49.825, 50.075]$ (GHz) & FL subcarrier ($S=5$)  \\
            GW antenna diameter \cite{ESA_HITEC} & $6.8$ m \\
            Satellite Orbit							&$13^{\circ}$E (GEO)		\\
            GEO Rx antenna diameter \cite{Delamotte_TAES20} & $1.4$ m \\       
            Separation between $2$ GEO Rx-antennas \cite{Delamotte_TAES20} & $3$ m \\ 
            Miscellaneous losses \cite{Delamotte_TAES20} & $1$ dB \\			
			Beam Hardware-Power  							& $5$ W\\
			Beam Radiation Pattern 					&Provided by ESA \\
			Downlink Carrier Frequency							& 19.5 GHz\\
			User Link Bandwidth, $R_s$							& 250 MHz\\ 
			Noise Power	at Satellite and Users											& $-121.3$ and $-118.6$ dB\\
			\bottomrule
		\end{tabular}
		\vspace{-0.2cm}
		\end{table}

  	\begin{figure}[!t]
		\centering
		\includegraphics[width=70mm]{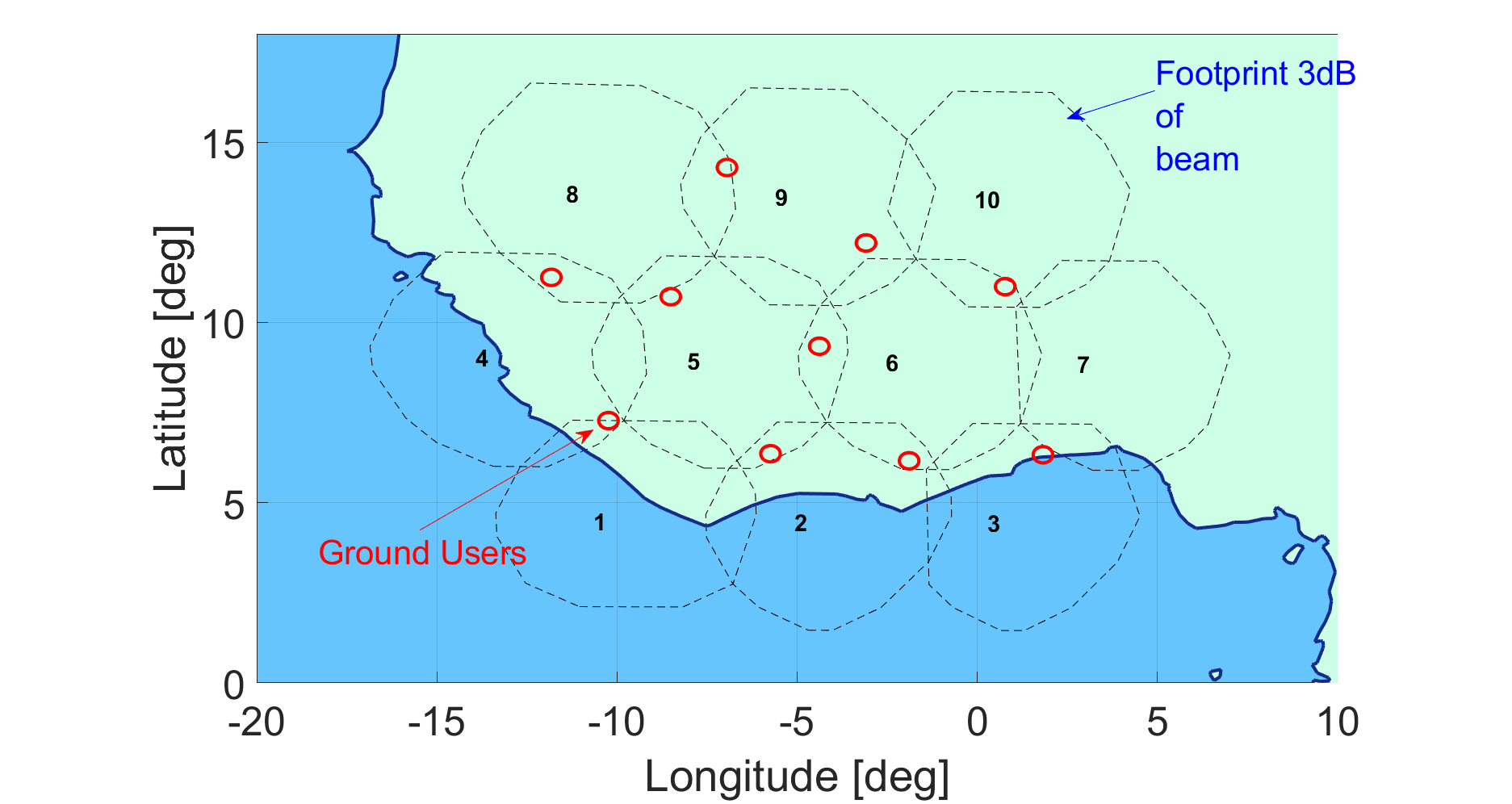}
		\vspace{-0.2cm}
  \captionsetup{font=footnotesize}
		\caption{Considered GEO multibeam footprint pattern with $N=10$.}
		\label{map_footprint}
			\vspace{-4mm}
	\end{figure}

  	We consider a GEO satellite system with $10$ spot beams serving $10$ users, i.e., $N=10$ and $K=10$, as shown in Fig.~\ref{map_footprint}. 
    Two GWAs located in Redu (Belgium), and Betzdorf (Luxembourg) with $(lat,lon)$ coordinates of $(50.002461, 5.148105)$ and $(49.692915, 6.327135)$ are assumed. 
    Other setting parameters are summarized in Table \ref{tab:simpara}. In addition, the efficiency factors of all antennas and HPAs are set at $60 \%$ and $\mb{A}=\mb{I}_{N \times N}$.
    
      \begin{figure}[!t]
		\centering
		\includegraphics[width=80mm]{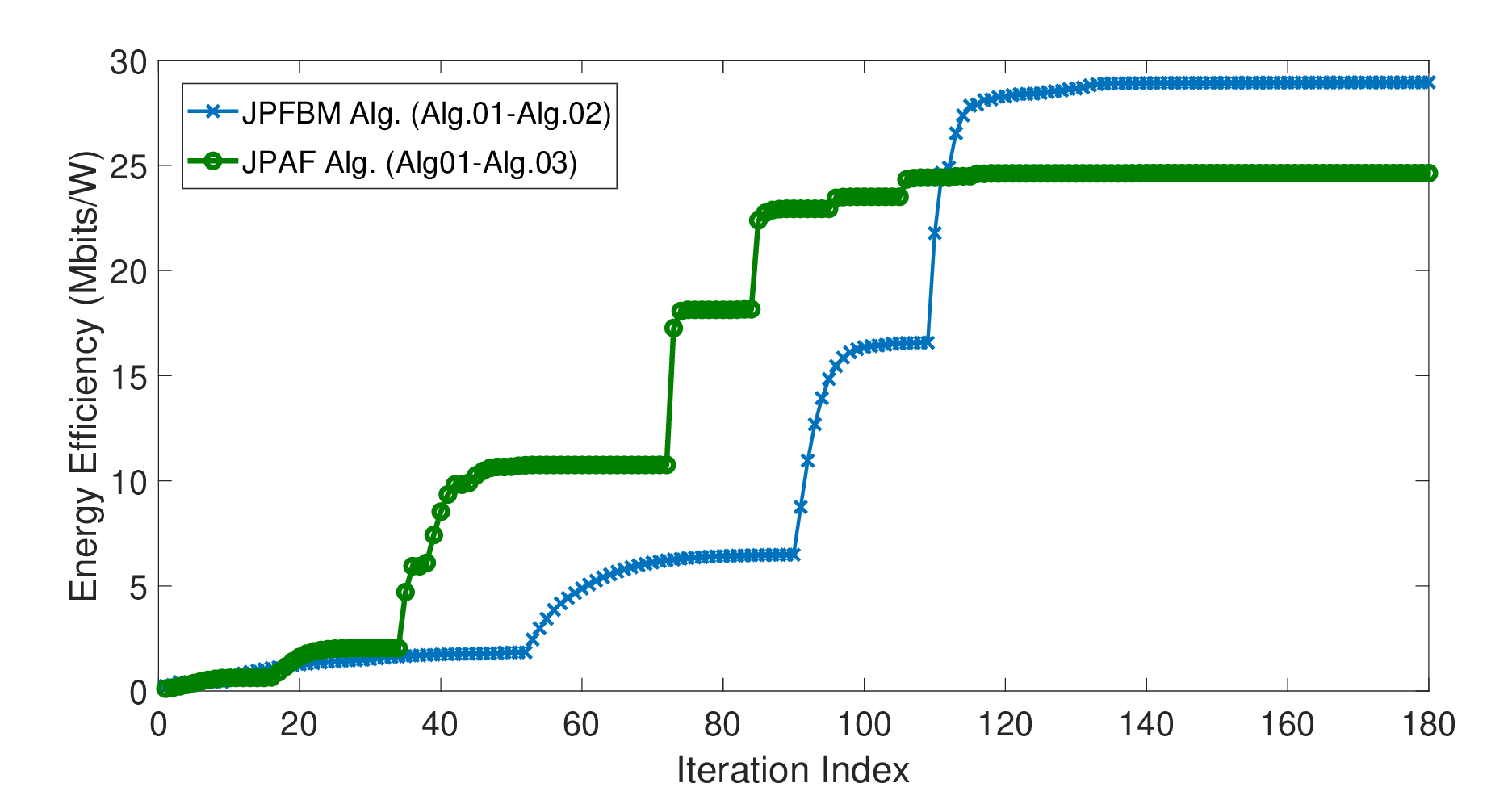}
		\vspace{-2mm}
		\captionsetup{font=footnotesize}
		\caption{System EE obtained by JPFBM and JPAF algorithms.}
		\label{Fig:Convg}
  \vspace{-6mm}
\end{figure}
     First, we investigate the convergence of our proposed algorithms by presenting the SEE results obtained by the JPFBM and JPAF frameworks over iterations in Fig.~\ref{Fig:Convg}.
     Here, $\bar{P}^{\sf{GWA}}_t = 15$~dBW and $\bar{P}^{\mr{Sat}}_n = 5$~dBW, 
     and we consider the total power consumption of the system by setting $\delta^{\sf{GWA}}=\delta^{\sf{Sa}}=1$. As observed, the SEEs for both approaches increase and plateau after around $100$ iterations, confirming the convergence of our proposed frameworks. Upon convergence, the JPFBM framework yields a higher SEE compared to the JPAF one. 

\begin{figure}[!t]
		\centering
		\includegraphics[width=80mm]{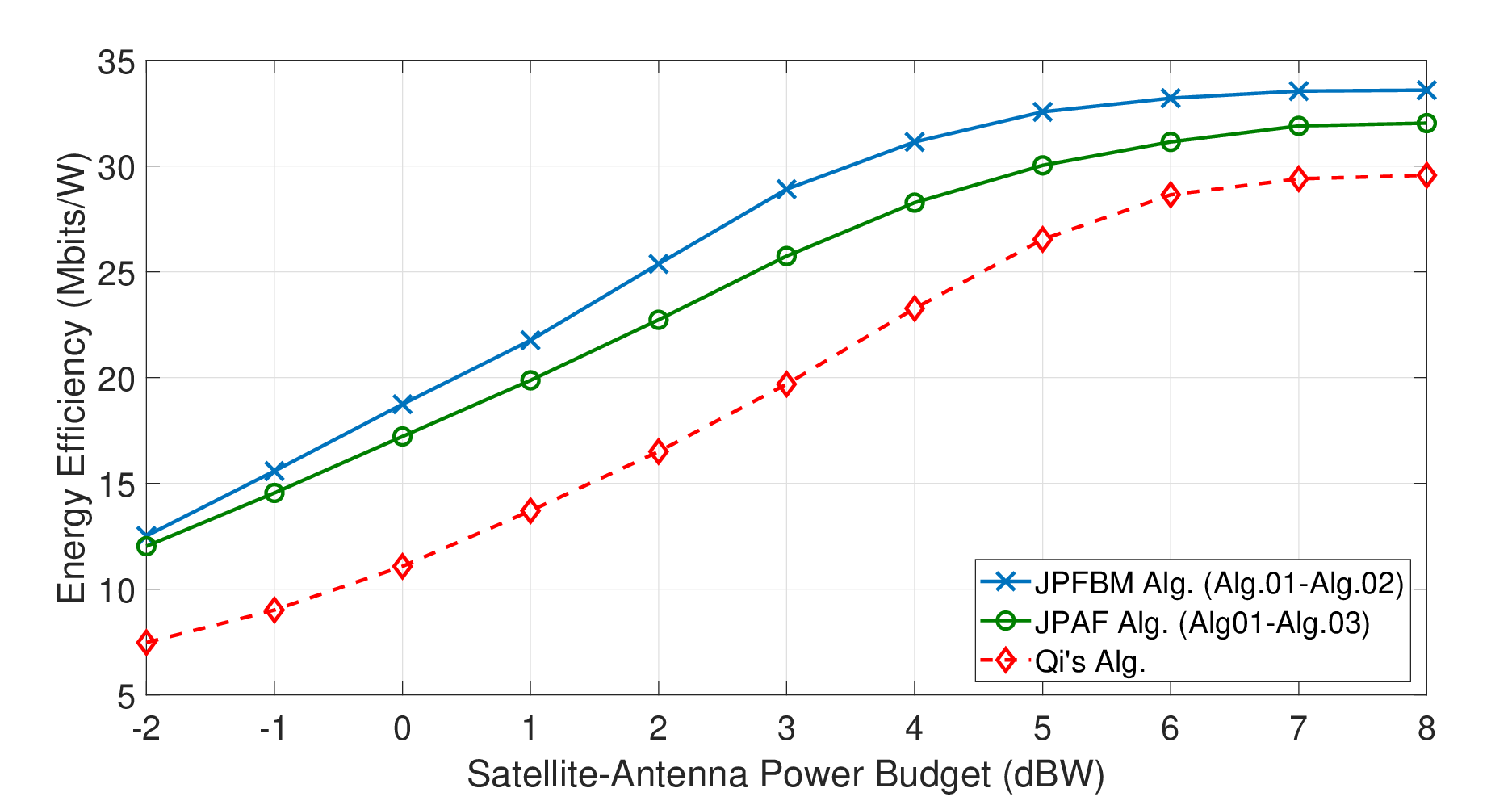}
		\vspace{-2mm}
		\captionsetup{font=footnotesize}
		\caption{The SEE versus the transmit power budget of each satellite antenna.}
		\label{SEE_SatTxPower}
  \vspace{-4mm}
\end{figure}
Next, Fig.~\ref{SEE_SatTxPower} depicts the SEE achieved by our proposed frameworks, as well as of Qi's method \cite{Qi_WCL2020}, with respect to varying values of $\bar{P}^{\sf{Sa}}_n$, the transmission power budget for each satellite antenna. 
Note that Qi's work only focuses on satellite power consumption in their SEE formula. 
To ensure a fair comparison, we set $\mb{B}=\left(\mb{F}^H\mb{F}\right)^{-1/2}$ and carry out simple manipulations to estimate the GW power consumption in this approach.
Here, $\bar{P}^{\sf{GWA}}_t = 15$~dBW. 
As expected, all three methods can achieve higher SEEs as $\bar{P}^{\sf{Sa}}_n$  increases. 
At the high regime of $\bar{P}^{\sf{Sa}}_n$, SEEs of these three tend to saturate due to the limitation in FL transmission. 
The figure also reveals that our proposed JPFBM and JPAF mechanisms surpass Qi's algorithm while JPFBM performs better than JPAF. 

\begin{figure}[!t]
		\centering
		\includegraphics[width=80mm]{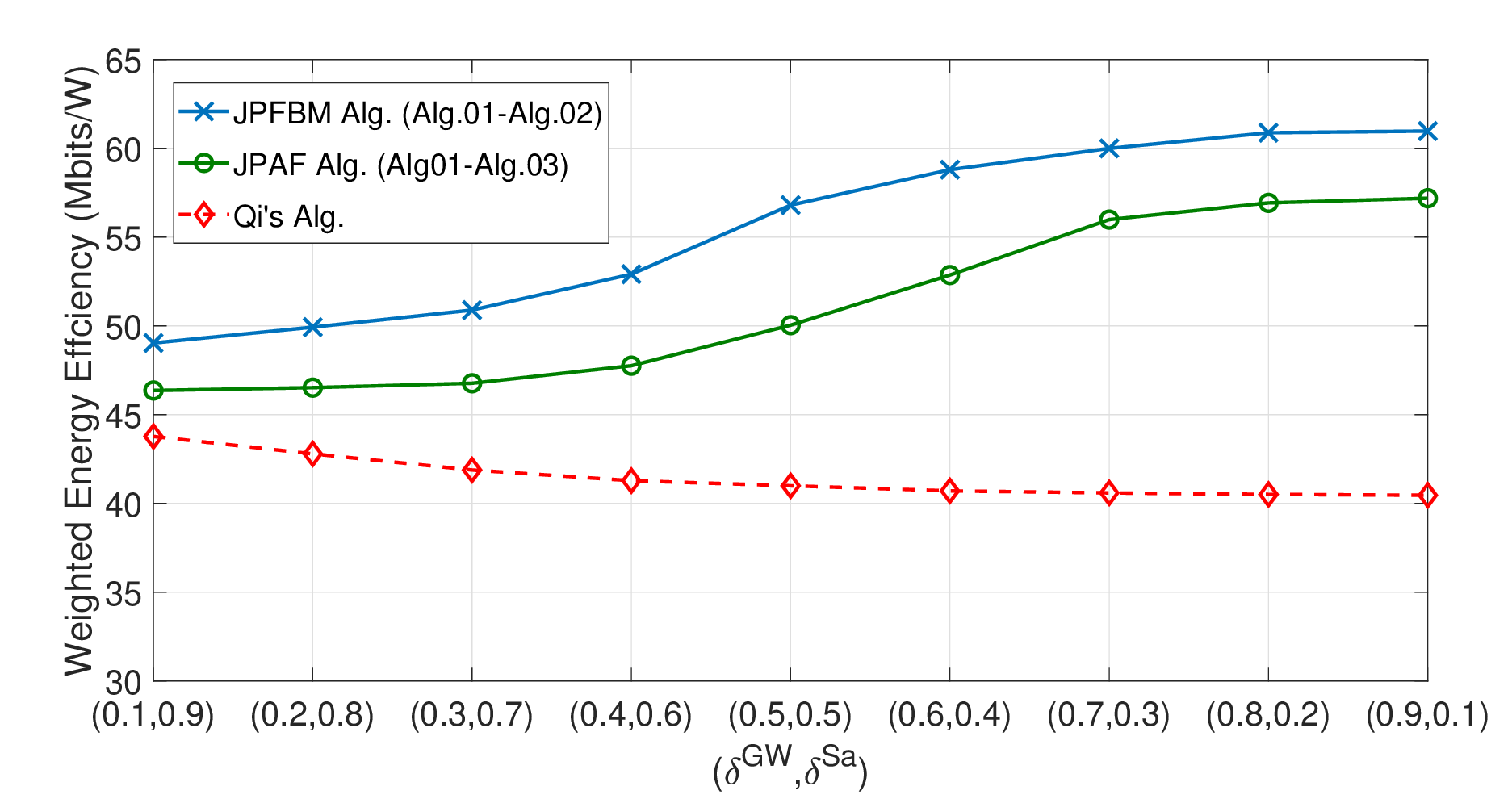}
		\vspace{-2mm}
		\captionsetup{font=footnotesize}
		\caption{The SWEE versus various values of $(\delta^{\sf{GWA}},\delta^{\sf{Sa}})$.}
		\label{SEE_delta}
  \vspace{-4mm}
\end{figure}

\begin{figure}[!t]
		\centering
		\includegraphics[width=80mm]{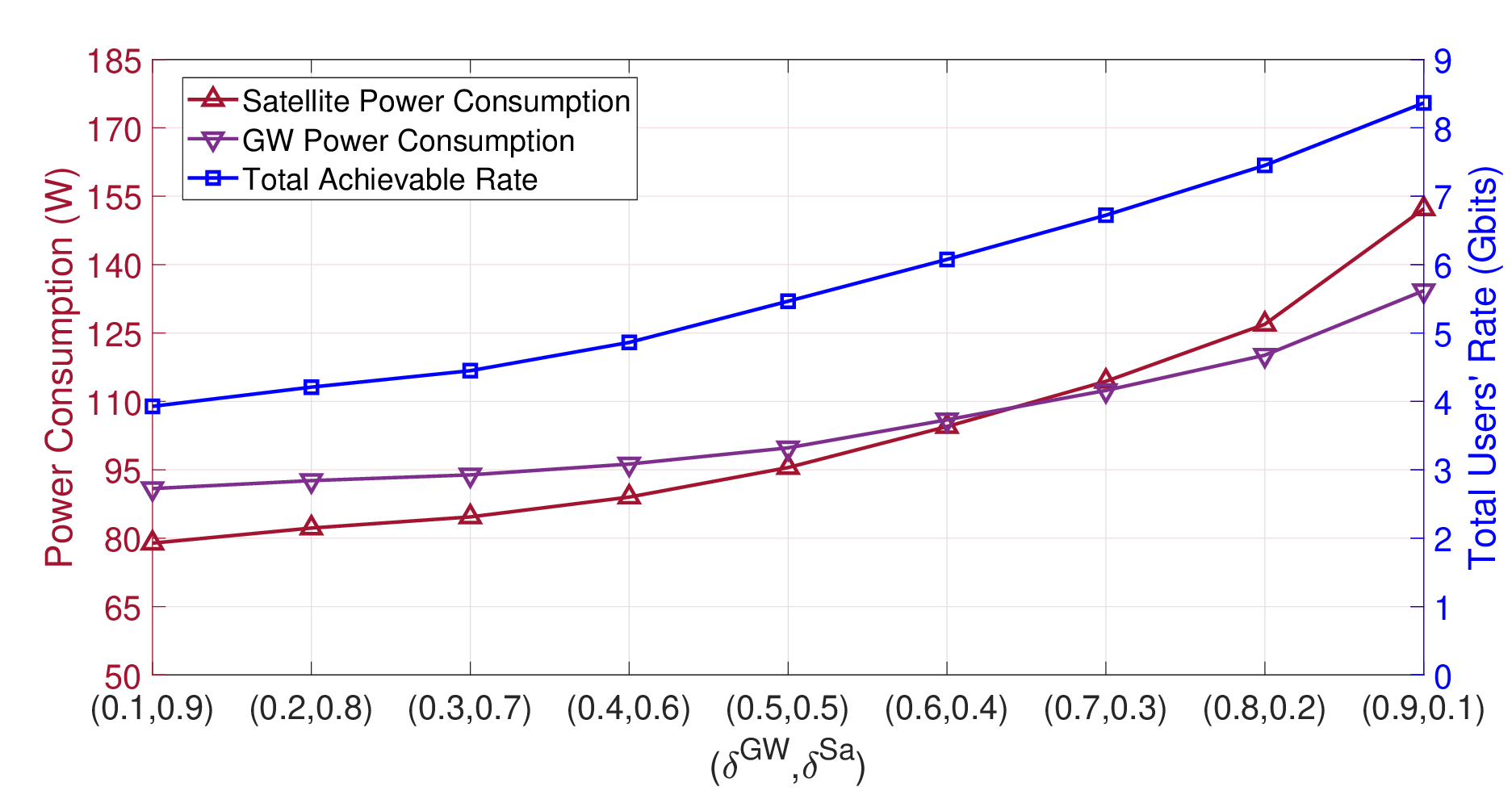}
		\vspace{-2mm}
		\captionsetup{font=footnotesize}
		\caption{The JPFBM sum rate and power consumption vs. $(\delta^{\sf{GWA}},\delta^{\sf{Sa}})$.}
		\label{Rate_Power_delta}
  \vspace{-4mm}
\end{figure}
Finally, Figs.~\ref{SEE_delta} and \ref{Rate_Power_delta} illustrate the variations in SWEE, achievable rate, and power consumption of JPFBM mechanism across different values of $(\delta^{\sf{GWA}},\delta^{\sf{Sa}})$.
In this simulation, we set $\bar{P}^{\sf{GWA}}_t = 15$~dBW, $\bar{P}^{\mr{Sat}}_n = 5$~dBW, vary $(\delta^{\sf{GWA}},\delta^{\sf{Sa}})$ such that $\delta^{\sf{GWA}}+\delta^{\sf{Sa}}=1$.
In Fig.~\ref{SEE_delta}, the SWEE of our proposed JPFBM and JPAF mechanisms increases while that of Qi's method decreases as $\delta^{\sf{GWA}}$ increases.
Once again, our proposed approaches outperform Qi's method across all power-weight configurations, and the JPFBM mechanism provides superior.
These results clearly emphasize the benefits of employing the jointly designed LP and FL-beam matching mechanism in the multi-GW, multi-beam SATCOM systems.
In Fig.~\ref{Rate_Power_delta}, one depicts that all rate and power consumption enlarge as $\delta^{\sf{GWA}}$ increases.
These outcomes shown in Figs.~\ref{SEE_delta} and \ref{Rate_Power_delta} suggest that the user links have a more significant impact on the network performance compared to the FLs. 

\section{Conclusion} \label{ccls}
\vspace{-1mm}
This paper presented a joint optimization framework for energy-efficient precoding and FL-beam matching design in multi-GW, multi-beam bent-pipe SATCOM systems which aims to maximize the SWEE. 
The technical designs were formulated as a non-convex sparsity problem. Two iterative efficient designs have been proposed to tackle these challenges by employing Dinkelbach and CS approach.
The simulation results showcased the effectiveness and superiority of the proposed JPFBM and JPAF frameworks over another benchmark. 

\vspace{-1mm}
\section*{Acknowledgment}
\vspace{-1mm}
This research was funded in whole by the Luxembourg National Research Fund (FNR), with three granted projects corresponding to three grant references
C23/IS/18073708/SENTRY (SENTRY project), C21/IS/16352790/ARMMONY (ARMMONY), and C19/IS/13696663/FLEXSAT (FlexSAT).
\vspace{-1mm}

\bibliographystyle{IEEEtran}
\bibliography{reference}

\end{document}